\documentstyle[11pt]{article}
\newlength{\bredde}
\def\slash#1{\settowidth{\bredde}{$#1$}\ifmmode\,\raisebox{.15ex}{/}
\hspace*{-\bredde} #1\else$\,\raisebox{.15ex}{/}\hspace*{-\bredde} #1$\fi}
\newcommand{\beq}{\begin{equation}}
\newcommand{\eeq}{\end{equation}}
\newcommand{\noi}{\vspace{12pt}\noindent}
\newcommand{\lG}{\raise.3ex\hbox{$\stackrel{\leftarrow}{G}$}}
\newcommand{\lU}{\raise.3ex\hbox{$\stackrel{\leftarrow}{U}$}}
\newcommand{\lP}{\raise.3ex\hbox{$\stackrel{\leftarrow}{{\cal P}}$}}
\newcommand{\leta}{\raise.3ex\hbox{$\stackrel{\leftarrow}{\eta}$}}
\newcommand{\lOmega}{\raise.3ex\hbox{$\stackrel{\leftarrow}{\Omega}$}}
\newcommand{\ldr}{\raise.3ex\hbox{$\stackrel{\leftarrow}{\delta^r}$}}

\def\beqn{\begin{eqnarray}}
\def\eeqn{\end{eqnarray}}

\def\gtwid{\raise.3ex\hbox{$>$\kern-.75em\lower1ex\hbox{$\sim$}}}
\def\ltwid{\raise.3ex\hbox{$<$\kern-.75em\lower1ex\hbox{$\sim$}}}

\def\la{\lambda}
\def\matA{{\mbox{\bf A}}}
\def\matB{{\mbox{\bf B}}}
\def\matC{{\mbox{\bf C}}}
\def\matD{{\mbox{\bf D}}}

\begin{document}
\topmargin -1.4cm
\oddsidemargin -0.8cm
\evensidemargin -0.8cm
\title{\Large{{\bf Microscopic Spectra of Dirac Operators and Finite-Volume
Partition Functions}}}

\vspace{1.5cm}

\author{~\\{\sc G. Akemann}\\Centre de Physique Th\'eorique CNRS\\
Case 907 Campus de Luminy\\F-13288 Marseille Cedex 9\\France\\~\\
{\sc P.H. Damgaard}\\
The Niels Bohr Institute\\ Blegdamsvej 17\\ DK-2100 Copenhagen {\O}\\
Denmark}
\date{} 
\maketitle
\vfill
\begin{abstract}Exact results from random matrix theory are used to
systematically analyse the relationship between microscopic Dirac
spectra and finite-volume partition functions. Results are presented
for the unitary ensemble, and the chiral analogs of the three classical
matrix ensembles: unitary, orthogonal and symplectic, all of which
describe universality classes of $SU(N_c)$ gauge theories with $N_f$ fermions
in different representations. Random matrix theory 
universality is reconsidered in this new light. 
\end{abstract}
\vfill
\begin{flushleft}
CPT-97/P.3571\\
NBI-HE-98-01 \\
hep-th/9801133
\end{flushleft}
\thispagestyle{empty}
\newpage

\section{Introduction}

\noi
One puzzle of the random matrix theory approach \cite{V,VZ,ADMN} to the
computation of Dirac operator spectra has been the case of one fermion
species. While the (massless) microscopic spectral densities associated
with the three chiral analogs of classical matrix ensembles have been
found to be different \cite{V}, the corresponding finite-volume partition
functions are all equal \cite{LS,SmV}\footnote{When considered in the
appropriate ``mesoscopic'' scaling region. Finite-volume partition
functions are always taken to be in this regime in what follows.}. This
means that all three microscopic spectral densities, despite being very
different, should lead to the same spectral sum rules of, for instance,
the kind ($\nu$ indicates the (positive) topological charge):
\beq
\left\langle \sum_n\! ~' \frac{1}{\la_n^2}\right\rangle_{\nu} ~=~
\frac{1}{4(\nu+1)}
\eeq
which indeed they do \cite{V}. This holds also in the case of
double-microscopic spectral densities, where fermion masses are kept
finite (and scaled with volume at the same rate as the eigenvalues)
\cite{D}.

\noi
As we shall show in this paper, there are in fact simultaneously two
reasons for why three different universality classes of gauge theories
can share the same finite-volume partition functions, have in common
an infinite set of spectral sum rules, and yet have very different
microscopic spectral densities. Let us first focus on the
chiral unitary ensemble, which corresponds to $SU(N_c)$ gauge theories
with $N_c\!\geq\! 3$ and $N_f$ fermions in the fundamental representation
of the gauge group. As was explained in ref. \cite{D1}, to compute the
(double-) microscopic spectral density $\rho_S^{(N_{f})}(\zeta;\mu_1,
\ldots, \mu_{N_{f}})$ corresponding to this random matrix ensemble, one
needs, in addition to the finite-volume partition 
${\cal Z}^{(N_{f})}(\mu_1,\ldots,\mu_{N_{f}})$ of the theory with $N_f$
flavors, also that of a theory with two additional fermion species (of
imaginary mass), 
${\cal Z}^{(N_{f}+2)}(\mu_1,\ldots,\mu_{N_{f}},i\zeta,i\zeta)$. So for
$N_f\!=\! 1$ we need also the partition function for $N_f\!=\! 3$ (and
the finite-volume partition functions corresponding to the two other
random matrix ensembles for $N_f\!=\! 3$ are {\em different}). In addition,
and this is one of the main points of this paper,
also the precise relationship between the (double-) microscopic spectral
densities and the corresponding finite-volume partition functions turn
out to be different.

\noi
One case which stands on a rather special footing is that of the (non-chiral)
unitary ensemble. It has been conjectured to be relevant for QCD-like theories
in $(2\!+\! 1)$ dimensions with an even number of fermions species. In
section 2 of this paper we begin by deriving the relations which allow us
to directly compute all double-microscopic spectral correlators from the
finite-volume partition functions of QCD$_3$ alone. We also show how the
universal limits of the orthogonal polynomials of the matrix model
formulation can be computed directly from a QCD$_3$ partition function.
In section 3 we turn
to the chiral unitary ensemble described above, where we do the  
extension of the analysis of ref. \cite{D1} to the case of arbitrary
topological charge $\nu$, and also here demonstrate how the universal
microscopic limit of the associated orthogonal polynomials are given
in terms of finite-volume partition functions. 
In section 4 we describe the extent to which we have been able
to derive analogous relations
for the two remaining categories of (chiral) random matrices:
the orthogonal and symplectic ensembles. We end in section 5 with our
conclusions and a discussion of the notion of matrix-model universality
\cite{ADMN,DN1,DN}, as seen in this new light. 

\section{The Unitary Ensemble}

\noi
The unitary ensemble has been conjectured to describe, in the microscopic
limit, the Dirac operator spectrum of $(2\!+\! 1)$-dimensional
$SU(N_c)$ gauge theories with
$N_c\!\geq\! 3$ and an {\em even} number of fermions $N_f$ \cite{VZ}.
The relevant partition function of that ensemble is, for massive fermions,
\beq
\tilde{\cal Z}^{(N_{f})}(m_1,\ldots,m_{N_{f}})
~=~ \int\! dM \prod_{f=1}^{N_{f}}{\det}\left(M + im_f\right)~ 
\mbox{e}^{-{N} \mbox{tr}\, V(M^2)} ~,
\label{ZUE}
\eeq
where the integration is over the Haar measure of 
hermitian $N\times N$ matrices $M$. The even potential $V(M^2)$
has been left unspecified, and we write it in general as
\beq
V(M^2) = \sum_{k\geq 1} \frac{g_{k}}{2k} M^{2k} ~.
\eeq

\noi
It is important to note that in order for this to possibly describe the 
microscopic limit of the above class of gauge theories, the even number 
of fermions have been regrouped into two sets of which one half is assigned the
same values as the other half, except for a minus sign, $i.e.$, the mass 
matrix can be taken to be of the form 
$$ {\mbox{\rm diag}}
(m_1,m_2,\ldots,m_{N_{f}/2},-m_1,-m_2,\ldots,-m_{N_{f}/2}) ~.
$$
With this assignment, the fermions can effectively be regrouped into 4-spinors,
the dynamics of which is reminiscent of the corresponding 
$(3\!+\! 1)$-dimensional gauge theories. The associated ``chiral symmetry'',
which really is a flavor symmetry, has been conjectured to break
spontaneously according to $U(N_f)\!\to\! U(N_f/2)\!\times\! U(N_f/2)$
(recall that $N_f$ is even) \cite{P,VZ}. An order parameter for this
symmetry breaking pattern is the absolute value of the chiral condensate
$\Sigma\!=\!\sum_i|\langle\bar{\psi}_i\psi_i\rangle|/N_f$. In matrix model
terminology a non-zero condensate translates into a non-zero spectral
density at the origin, $\rho(0)\!\neq\! 0$. For the underlying field theory,
this is simply a $(2\!+\! 1)$-dimensional generalization of the 
Banks-Casher relation between the chiral condensate and the spectral density
of the Dirac operator, evaluated at the origin.  

\noi
In terms of the eigenvalues $\lambda_i$ of the hermitian
matrix $M$ we have, ignoring all irrelevant overall factors,
\beq
\tilde{\cal Z}^{(N_f)}(m_1,\ldots,m_{N_{f}}) ~=~
\int_{-\infty}^{\infty}\! \prod_{i=1}^N \left(d\lambda_i
\prod_{f=1}^{N_f/2}(\lambda_i^2 + m_f^2)~
\mbox{e}^{-NV(\lambda_i^2)}\right)\left|{\det}_{ij}
\lambda_j^{i-1}\right|^2 ~.
\eeq
The imaginary part arising from the term $im_f$ in eq. (\ref{ZUE}) has
disappeared due to the pairwise grouping of masses discussed above, and
the final expression is, except for the ``fermion determinant'', a conventional
eigenvalue integration in the unitary ensemble with arbitrary even potential. 
The massive
spectral correlators of this ensemble have recently been derived in the
double-microscopic limit where masses and eigenvalues are considered on
the same scale of magnification around the origin \cite{DN1}, and have
been proved to be universal in the random matrix theory context \cite{DN1}.

\noi
We are now in a position to make use of a nice result 
from ref. \cite{ZJ} (see also ref. \cite{MM}), 
where the spectral two-point function, the kernel
$K_N(x,y)$, was expressed in terms of a slightly modified random matrix
integral. By definition,
\beq
K^{(N_{f},\nu)}_N(x,x';m_1,\ldots,m_{N_{f}}) ~=~ 
e^{-\frac{N}{2}(V(x^2)+V(x'^2))}
\prod_{f}^{N_f/2}\sqrt{(x^2+m_f^2)(x'^2+m_f^2)}
\sum_{i=0}^{N-1} P_i(x)P_i(x') ~,
\eeq 
where $P_n(x)$ is an $n$th order polynomial orthogonal with respect to
the even weight function
\beq
w(x) ~=~ \prod_{f=1}^{N_f/2}(x^2 + m_f^2)~
\mbox{e}^{-NV(x^2)} ~,\label{w}
\eeq
and this polynomial is hence also a function of all masses $m_i$.
For convenience we consider here the normalization where the polynomials
are chosen orthonormal on the real line. By making use of the orthonomality
one can write the kernel in the form of a matrix integral:
\begin{eqnarray}
K^{(N_{f})}_N(x,x';m_1,\ldots,m_{N_{f}})\! &=& \!
\frac{e^{-\frac{N}{2}(V(x^2)+V(x'^2))}
\prod_{f}^{N_f/2}\sqrt{(x^2+m_f^2)(x'^2+m_f^2)}}{
{\tilde{\cal Z}}^{(N_{f})}(m_1,\ldots,m_{N_{f}})} ~\times \cr
&& \!
\int_{-\infty}^{\infty}\! \prod_{i=1}^{N-1}\!\!\left(\!d\lambda_i 
(\la_i-x)(\la_i-x')\!\!\prod_{f=1}^{N_{f}/2}(\lambda_i^2 + m_f^2)
\mbox{e}^{-NV(\lambda_i^2)}\!\right)\!\left|{\det}_{ij}
\lambda_j^{i-1}\right|^2 \! . 
\end{eqnarray}
Although the last matrix integral is over $(N\!-\! 1)$ eigenvalues only, 
this distinction becomes irrelevant in the large-$N$ limit, which we will
consider below. This gives us a very convenient expression for the spectral
two-point function in the large-$N$ limit:
\begin{eqnarray}
K^{(N_{f})}_N(x,x';m_1,\ldots,m_{N_{f}}) &=&
e^{-\frac{N}{2}(V(x^2)+V(x'^2))}
\prod_{f}^{N_f/2}\sqrt{(x^2+m_f^2)(x'^2+m_f^2)}\cr
&&\times ~\frac{
\tilde{\cal Z}^{(N_{f}+2)}(m_1,\ldots,m_{N_{f}},ix,ix')}{
\tilde{\cal Z}^{(N_{f})}(m_1,\ldots,m_{N_{f}})} ~,\label{KtildeZUE}
\end{eqnarray}
The partition functions are of course symmetric in the
mass entries, and the position of the additional fermion masses in
the argument list is therefore immaterial.

\noi
In writing the kernel in the form (\ref{KtildeZUE}) we have made use of
the observation that the insertion of the additional factor of
$(\la_i-x)(\la_i-x')$ in the matrix integral can be viewed as
considering a theory with two additional fermion species, of imaginary
masses (cf. eq. (\ref{ZUE})). Except for the shown prefactors, we have
thus expressed the two-point spectral correlation function in terms of
the ratio of two unitary-ensemble matrix model partition functions.
All higher-order correlation functions are then also manifestly expressed
in terms of these partition function through the usual factorization 
relation of the large-$N$ limit. Finally, also the conventional 
(macroscopic) spectral density is expressed in such a manner, 
now in terms of two additional fermion species degenerate in (imaginary) 
masses:
\beq
\rho^{(N_{f})}(x;m_1,\ldots,m_{N_{f}}) ~=~
\lim_{N\to\infty} K_N^{(N_{f})}(x,x;m_1,\ldots,m_{N_{f}}) ~.\label{rhoUE}
\eeq  

\noi
The above results hold in all generality in the planar limit.
As in ref. \cite{D1}, we now consider specifically the double-microscopic 
limit in which 
\beq
\zeta \equiv \pi\rho(0)Nx~~~~~ {\mbox{\rm and}}~~~~~ 
\mu_i \equiv \pi\rho(0)Nm_i
\eeq
are kept fixed in the limit 
$N\!\to\!\infty$. If we identify $\Sigma\!=\! \pi\rho(0)$, this is precisely
the mesoscopic scaling region of the finite volume partition functions of
the underlying gauge theories. In ref. \cite{VZ} these partition functions
were argued to be describable in terms of a very simple chiral Lagrangian,
in analogy with the Leutwyler-Smilga analysis in $(3\!+\! 1)$ dimensions:
\beq
{\cal Z}^{(N_{f})}(\mu_1,\ldots,\mu_{N_{f}}) 
~=~ \int dU \exp[Tr({\cal M}U\Gamma_5U^{\dagger})] ~.
\label{qcd3}
\eeq
Here ${\cal M}$ is the mass matrix discussed above, rescaled by $N\Sigma$:
\beq
{\cal M} ~=~ {\mbox{\rm diag}}
(\mu_1,\mu_2,\ldots,\mu_{N_{f}/2},-\mu_1,-\mu_2,\ldots,-\mu_{N_{f}/2}) ~,
\eeq
and $\Gamma_5\!\equiv\!
({\mbox{\bf 1}}_{N_{f}/2},-{\mbox{\bf 1}}_{N_{f}/2})$ where {\bf 1}$_{
N_{f}/2}$ is an $(N_f/2)\!\times\!(N_f/2)$ unit matrix. The above
partition function can be explicitly evaluated by making use of the
Harish-Chandra--Itsykson-Zuber integral \cite{HCIZ}. The result is \cite{DN1}:
\beq
{\cal Z}^{(N_{f})}({\cal M}) ~=~ \frac{\det\left(
\begin{array}{ll}
\matA(\{\mu_i\}) & \matA(\{-\mu_i\})\\
\matA(\{-\mu_i\})  & \matA(\{\mu_i\})
\end{array}\right)}{\Delta({\cal M})} ~,\label{ZQCD3}
\eeq 
where the $(N_f/2)\!\times\!(N_f/2)$ matrix
$\matA(\{\mu_i\})$ is defined by 
\beq
\mbox{A}_{ij} ~\equiv~ (\mu_i)^{j-1} e^{\mu_i} ~. \label{matAdef}
\eeq
The denominator is given by the Vandermonde determinant of
rescaled masses:
\beq
\Delta({\cal M}) ~=~ \prod_{i<j}^{N_{f}}(\mu_i-\mu_j) ~.
\eeq
In the mesoscopic scaling limit the matrix model partition functions
${\tilde{\cal{Z}}}^{(N_f)}(\mu_1,\ldots,\mu_{N_{f}})$ should equal 
the field theory
partition function ${\cal Z}^{(N_f)}(\mu_1,\ldots,\mu_{N_{f}})$ up to
an irrelevant (mass-independent) normalization factor. Furthermore, in
this scaling regime the prefactor of $\exp[-(N/2)(V(x^2)+V(x'^2))]$ in
the expression for the kernel (\ref{KtildeZUE}) becomes replaced by unity.
For the kernel this leads us to the following master formula: 
\beq
K_S^{(N_{f})}(\zeta,\zeta';\mu_1,\ldots,\mu_{N_{f}}) ~=~
C \prod_{f}^{N_{f}/2}\sqrt{(\zeta^2+\mu_f^2)(\zeta'^2+\mu_f^2)}~\frac{
{\cal Z}^{(N_{f}+2)}(\mu_1,\ldots,\mu_{N_{f}},i\zeta,i\zeta')}{
{\cal Z}^{(N_{f})}(\mu_1,\ldots,\mu_{N_{f}})} ~.\label{mfUE}
\eeq
Similarly, the double-microscopic spectral density becomes
\beq
\rho_S^{(N_{f})}(\zeta;\mu_1,\ldots,\mu_{N_{f}}) ~=~
C \prod_{f}^{N_{f}/2}(\zeta^2+\mu_f^2)~\frac{
{\cal Z}^{(N_{f}+2)}(\mu_1,\ldots,\mu_{N_{f}},i\zeta,i\zeta)}{
{\cal Z}^{(N_{f})}(\mu_1,\ldots,\mu_{N_{f}})} ~,\label{mfrhoUE}
\eeq
and all other microscopic correlators are given by
\beq
\rho_S^{(N_{f})}(\zeta_1,\ldots,\zeta_n;\mu_1,\ldots,\mu_{N_{f}}) ~=~
\det_{a,b} K_S^{(N_{f})}(\zeta_a,\zeta_b;\mu_1,\ldots,\mu_{N_{f}}) ~.
\label{correlUE}
\eeq
In these expressions there is still one overall normalization factor, $C$,
which remains to be fixed. As discussed in ref. \cite{D1}, the simplest 
way to fix it is to make use of the matching condition between the
microscopic spectral density and macroscopic spectral density,
\beq
\lim_{\zeta\to\infty}\rho_S^{(N_{f})}(\zeta;\mu_1,\ldots,\mu_{N_{f}}) 
~=~ \rho(0) ~=~ \frac{1}{\pi} ~,\label{match}
\eeq
where we have inserted the conventional normalization \cite{V} (this
conveniently makes $\mu_i\!=\! Nm_i$ and $\zeta_i\!=\! Nz_i$).

\noi
To illustrate the power of the master formula (\ref{mfUE}), consider the
simplest case of quenched fermions, which formally corresponds to $N_f\!=\!0$.
In this case the origin $\la\!=\!0$ is not singled out, and the kernel
$K_S(\zeta,\zeta')$ should reduce to the famous sine-kernel of the unitary
ensemble. Indeed, the relevant partition function for two flavors in this 
case reads, from eq. (\ref{ZQCD3}),
\beq
{\cal Z}(\mu_1,\mu_2) ~=~ \frac{2\sinh(\mu_1-\mu_2)}{\mu_1-\mu_2} ~,
\eeq 
while the finite-volume partition function for $N_f\!=\!0$ is a trivial
constant, which we set to unity. The kernel is therefore given by
\beq
K_S(\zeta,\zeta') ~=~ C~ \frac{2\sin(\zeta-\zeta')}{\zeta-\zeta'} ~,
\eeq
and the microscopic spectral density, as expected, becomes just a constant:
\beq
\rho_S^{(0)}(\zeta) ~=~ 2C ~.
\eeq
The matching condition (\ref{match}) hence gives $C=1/(2\pi)$, and thus
\beq
K_S(\zeta,\zeta') ~=~ \frac{1}{\pi}\frac{\sin(\zeta-\zeta')}{\zeta-\zeta'} ~,
\eeq
- a novel derivation of the sine-kernel.

\noi
Because we know the analytical form of the finite-volume partition function
for any number $N_f$ of (massive) fermion species, we can immediately
write down the general expressions for all double-microscopic spectral
correlators. From eq. (\ref{ZQCD3}) it follows that
\beq
K_S^{(N_f)}(\zeta_1,\zeta_2;\mu_1,\ldots,\mu_{N_{f}}) = 
\frac{-iC}{(\zeta_1-\zeta_2)
\prod_{f}^{N_f/2}\!\!\sqrt{(\zeta_1^2+\mu_f^2)(\zeta_2^2+\mu_f^2)}}
\frac{\det\left(\!\!
\begin{array}{ll}
\matB(\{\mu_i,\zeta_1\}) & \matB(\{-\mu_i,-\zeta_1\})\\
\matB(\{-\mu_i,\zeta_2\})  & \matB(\{\mu_i,-\zeta_2\})
\end{array}\!\!\!\right)}{\det\left(
\begin{array}{ll}
\matA(\{\mu_i\}) & \matA(\{-\mu_i\})\\
\matA(\{-\mu_i\})  & \matA(\{\mu_i\})
\end{array}\right)} 
\label{KgenUE}
\eeq
where $\matA(\{\mu_i\})$ is the $(N_f/2)\!\times\!(N_f/2)$
matrix defined previously in eq. (\ref{matAdef}), and 
$\matB(\{\mu_i,\zeta_i\})$ is an
$(N_f/2\!+\!1)\!\times\!(N_f/2\!+\!1)$ matrix defined by
\begin{eqnarray}
B_{kl} &=& A_{kl} ~~~~~~~~~~~~~~~~~~~{\mbox{\rm for}}~~ 1 \leq k \leq 
\frac{N_{f}}{2}~~;~~ 1 \leq l \leq 
\frac{N_{f}}{2}+1 \cr
B_{kl} &=& (i\zeta_{1,2})^{l-1} e^{i\zeta_{1,2}} ~~~~~
{\mbox{\rm for}}~~ k = \frac{N_{f}}{2} + 1~~;~~ 1 \leq l \leq 
\frac{N_{f}}{2}+1 ~,
\end{eqnarray}
where in the last line the entry is either $\zeta_1$ or $\zeta_2$, as
indicated explicitly in eq. (\ref{KgenUE}).

\noi
It similarly follows from eq. (\ref{mfrhoUE}) that the double-microscopic 
spectral density is 
\beq
\rho_S^{(N_f)}(\zeta;\mu_1,\ldots,\mu_{N_{f}}) ~=~ \frac{-C}{
\prod_{f}^{N_{f}/2}(\zeta_1^2+\mu_f^2)}
\frac{\det\left(
\begin{array}{ll}
\matB(\{\mu_i,\zeta\}) & \matB(\{-\mu_i,-\zeta\})\\
\matC(\{-\mu_i,\zeta\})  & \tilde{\matC}(\{\mu_i,-\zeta\})
\end{array}\right)}{\det\left(
\begin{array}{ll}
\matA(\{\mu_i\}) & \matA(\{-\mu_i\})\\
\matA(\{-\mu_i\})  & \matA(\{\mu_i\})
\end{array}\right)} ~, \label{rhogenUE}
\eeq 
where $\matC(\{\mu_i,\zeta\})$ and $\tilde{\matC}(\{\mu_i,\zeta\})$ are 
$(N_f/2\!+\!1)\!\times\!(N_f/2\!+\!1)$ matrices defined by
\begin{eqnarray}
C_{kl} ~=~ \tilde{C}_{kl} ~=~ A_{kl} ~~~~~~~~{\mbox{\rm for}}~~ 1 \leq k \leq 
\frac{N_{f}}{2}~~;~~ 1 \leq l \leq 
\frac{N_{f}}{2}+1 ~, \cr
C_{kl} ~=~ -\tilde{C}_{kl} ~=~ (i\zeta)^{l-2} e^{i\zeta}(i\zeta + l - 1) 
~~~~~~~~{\mbox{\rm for}}~~ k = \frac{N_{f}}{2} + 1 ~~;~~ 1 \leq l \leq 
\frac{N_{f}}{2}+1 ~.
\end{eqnarray}
All other double-microscopic spectral correlators are then also known 
explicitly, using the relation (\ref{correlUE}).

\noi
It remains to fix the overall constant $C$ in this general case. We again 
do it by the matching condition (\ref{match}). This gives 
$$
C = \frac{1}{2\pi}~.
$$
All double-microscopic spectral correlators, for any even value of $N_f$,
are then completely determined. Recently these double-microscopic spectral
correlators were evaluated by means of random matrix theory, and 
the universality of the result in that framework was established 
\cite{DN1}. The above expressions for the same quantities are more compact 
and convenient. We have explicitly checked in some special cases that the 
results of ref. \cite{DN1} agree with those presented here.

\noi
It is interesting to note that not only can we compute the microscopic
spectral correlators directly from the corresponding finite-volume
partition functions, we can also derive the universal double-microscopic
limits of the orthogonal polynomials from these partition functions. This
despite of the fact that these orthogonal polynomials seem to have no
clear interpretation in field theory language. 

\noi
To derive expressions for the orthogonal polynomials, we make use of
a convenient representation of these polynomials in terms of matrix
integrals (see, $e.g.$, ref. \cite{GM}):
\beq
P_{2n}^{(N_f)}(\la;m_1,\ldots,m_{N_{f}}) ~=~ 
\frac{1}{\tilde{\cal Z}_{2n}^{(N_f)}
(m_1,\ldots,m_{N_{f}})}\int_{-\infty}^{\infty}
\prod_{i=1}^{2n}[d\la_i w(\la_i)]\left|{\det}_{ij}\lambda_j^{i-1}\right|^2 
\prod_{i=1}^{2n}(\la - \la_i) \label{pol}
\eeq
Here,
\beq
\tilde{\cal Z}_{2n}^{(N_f)}(m_1,\ldots,m_{N_{f}}) 
~=~ \int_{-\infty}^{\infty}\prod_i^{2n}[d\la_i w(\la_i)]
\left|{\det}_{ij}\lambda_j^{i-1}\right|^2 ~,
\eeq
and the measure factor $w(\la_i)$ is that of eq. (\ref{w}).
The relation (\ref{pol}) is readily verified by noting that it yields the 
required orthogonality relation. It also follows from (\ref{pol}) that
the normalization corresponds to monic polynomials, $i.e.$,
$P_n(\la) \!=\! \la^n + \ldots$.

\noi
One sees that for $n,N \to \infty$, with $t\equiv 2n/N$ fixed,  
eq. (\ref{pol}) becomes a relation
between the orthogonal polynomials and (matrix model) partition functions,
now involving an {\em odd} number of fermion species $N_f\!+\!1$ 
(the additional fermion having mass $i\la$). Let us now consider the 
double-microscopic scaling limit,
and take $t\!=\! 1$. The relation (\ref{pol}) then gives 
\beq
P_{2n}^{(N_{f})}(\zeta;\mu_1,\ldots,\mu_{N_{f}}) ~=~ C_1 
\frac{{\cal Z}^{(N_{f}+1)}(\mu_1,\ldots,\mu_{N_{f}},i\zeta)}
{{\cal Z}^{(N_{f})}(\mu_1,\ldots,\mu_{N_{f}})} ~, \label{polz}
\eeq
where we again have replaced the matrix model partition functions by those
of the finite-volume field theories (at the cost of introducing one
overall proportionality constant $C_1$).

\noi
We have indicated the relation for the even polynomials only. This is 
actually all that is required, since we can construct the odd polynomials
from the following relation \cite{ADMN}:
\beq
P_{2n+1}^{(N_{f})}(\la;m_1,\ldots,m_{N_{f}}) ~=~ 
\frac{\tilde{P}_{2n+2}^{(N_{f})}(\la;m_1,\ldots,m_{N_{f}}) - 
\tilde{P}_{2n}^{(N_{f})}(\la;m_1,\ldots,m_{N_{f}})}{\la} ~,
\label{evenodd}
\eeq
where the even polynomials 
$\tilde{P}_{2n}^{(N_{f})}(\la;m_1,\ldots,m_{N_{f}})$ are 
those of eq. (\ref{pol}),
but in a different normalization: 
$\tilde{P}_{2n}^{(N_{f})}(0;m_1,\ldots,m_{N_{f}}) = 1$.

\noi
The distinction between odd and even polynomials in the manner shown
above actually has a consistent interpretation in terms of the finite-volume
partition functions from field theory. Verbaarschot and Zahed \cite{VZ}
have given these finite-volume partition functions in the form of 
integrals:
\beq
{\cal Z}^{(N_{f}+1)}({\cal M}) ~=~ 
\int\! dU~ \cosh[Tr({\cal M}U\Gamma U^{\dagger})]\label{zeven}
\eeq
for $N$ even, and
\beq
{\cal Z}^{(N_{f}+1)}({\cal M}) ~=~ 
\int\! dU~ \sinh[Tr({\cal M}U\Gamma U^{\dagger})]\label{zodd}
\eeq
for $N$ odd (where $N$ denotes the three-volume). 
Here ${\cal M}$ is the  mass matrix already rescaled by $N\Sigma$, 
whose diagonal form is
\beq
{\cal M} ~=~ {\mbox{\rm diag}}
(\mu,\mu_1,\mu_2,\ldots,\mu_{N_{f}/2},-\mu_1,-\mu_2,\ldots,-\mu_{N_{f}/2}) ~,
\label{Modd}
\eeq
and $\Gamma\!\equiv\! {\mbox{\rm diag}}({\mbox{\bf 1}}_{N_{f}/2+1},
-{\mbox{\bf 1}}_{N_{f}/2})$. We can choose to use just the even-$N$
partition functions (\ref{zeven}) to get the even polynomials (\ref{polz}),
and then derive the odd polynomials from the relation (\ref{evenodd}).
Alternatively, we can use the odd-$N$ partition function directly (the analogue
of the formula (\ref{pol}) for odd polynomials). The result
will be the same, as follows by expanding the partition function
(\ref{zeven}) for $N+2$ in terms of the partition function for $N$
and then comparing with eq. (\ref{evenodd}). In fact, just from the
relations (\ref{evenodd}) and (\ref{polz}) it follows that if the even-$N$
partition function is given by (\ref{zeven}), then the odd-$N$ partition
function must necessarily be given by (\ref{zodd}).

\noi
Both of the above partition functions are given in terms of integrals
of the $SU(N_f\!+\!1)$-invariant Haar measure $dU$. These integrals can
actually be evaluated rather easily by again making use of the
Harish-Chandra--Itsykson-Zuber integration formula \cite{HCIZ}. The
case of one flavor of course stands on a special footing, and there
one has, trivially,
\beq
{\cal Z}^{(1)}(\mu) ~=~ \cosh(\mu) 
\label{Zcosh}
\eeq
for $N$ even, and
\beq
{\cal Z}^{(1)}(\mu) ~=~ \sinh(\mu) 
\label{Zsinh}
\eeq
for $N$ odd. Here $\mu$ is the rescaled mass: $\mu\!=\! m\Sigma N$, and
we have dropped all irrelevant ($\mu$-independent) overall constants. For
higher values of $N_f+1$ we find
\beq
{\cal Z}^{(N_{f}+1)}(\{\mu;\mu_i\}) ~=~ \frac{1}{\Delta({\cal M})}\left[
\det\matD(\{\mu;\mu_i\}) + (-1)^{N+N_f/2}\det\matD(\{-\mu;-\mu_i\}) \right] ~,
\label{ZQCD3oddNf}
\eeq 
where $\Delta({\cal M})$ is the the Vandermonde determinant of ${\cal M}$
eq. (\ref{Modd}) and the 
additional sign $(-1)^{N_f/2}$ originates from $\Delta(-{\cal M})$.
The matrix $\matD$ 
is of size $(N_f\!+\!1)\!\times\!(N_f\!+\!1)$ and is given by
\begin{eqnarray}
D_{1j} &=& \mu^{j-1}e^{\mu}~~~~~~~~~~~~~~~~~~~~~~~~
{\mbox{\rm for}}~~ 1 \leq j \leq 
\frac{N_{f}}{2}+1 \cr
D_{1j} &=& (-\mu)^{j-\frac{N_{f}}{2}-2}e^{-\mu}~~~~~~~~~~~~{\mbox{\rm for}}~~ 
\frac{N_{f}}{2}+2 \leq j \leq N_{f}+1 \cr
D_{ij} &=& \mu_{i-1}^{j-1}e^{\mu_{i-1}}~~~~~~~~~~~~~~~~~~~~
{\mbox{\rm for}}~~ 2 \leq i \leq \frac{N_{f}}{2}+1 ~~;~~ 
1 \leq j \leq \frac{N_{f}}{2}+1 \cr
D_{ij} &=& (-\mu_{i-1})^{j-\frac{N_{f}}{2}-2}e^{-\mu_{i-1}}~~~~
{\mbox{\rm for}}~~ 2 \leq i \leq \frac{N_{f}}{2}+1 ~~;~~ 
\frac{N_{f}}{2}+2 \leq j \leq N_{f}+1 \cr
D_{ij} &=& \mu_{i-1}^{j-1}e^{\mu_{i-1}}~~~~~~{\mbox{\rm for}}~~ 
\frac{N_{f}}{2}+2 \leq i \leq N_{f}+1 ~~;~~ 
1 \leq j \leq \frac{N_{f}}{2}+1 \cr
D_{ij} &=& (-\mu_{i-1})^{j-\frac{N_{f}}{2}-2}e^{-\mu_{i-1}}~~~~~~~~~~~
{\mbox{\rm for}}~~ \frac{N_{f}}{2}+2 \leq i \leq N_{f}+1 ~~;~~ 
\frac{N_{f}}{2}+2 \leq j \leq N_{f}+1 ~. 
\label{matDdef}
\end{eqnarray}

\noi
We are now ready to state the results. For the polynomials we get 
\beq
P_{N}^{(N_{f})}(\zeta;\mu_1,\ldots,\mu_{N_{f}}) ~=~ 
\frac{C_1(-1)^{N_f/2}}{\prod_f^{N_f/2}(\zeta^2+\mu_f^2)}
\frac{\det\matD(\{i\zeta;\mu_i\}) + (-1)^{N+N_f/2}
\det\matD(\{-i\zeta;-\mu_i\})}
{\det\left(
\begin{array}{ll}
\matA(\{\mu_i\}) & \matA(\{-\mu_i\})\\
\matA(\{-\mu_i\})  & \matA(\{\mu_i\})
\end{array}\right)}  
\eeq
where in matrix $\matD$ eq. (\ref{matDdef}) $\mu$ has been replaced by $i\zeta$
and $\matA$ is the matrix defined in eq. (\ref{matAdef}). The normalization 
constant $C_1$ remains undetermined, but can of course in any case be chosen
at will. In \cite{DN1} universal expressions
for the same polynomials have been derived from random matrix theory.
In contrast to the situation for the correlation functions eqs. (\ref{KgenUE})
and (\ref{rhogenUE})
the expression for the polynomials obtained here are less compact that in
\cite{DN1}. We have checked explicitly in some special cases that both results
agree.

\noi
To illustrate the simplicity of the derivation presented here,
let us look at the special case of quenched fermions $N_f=0$.
{}From eqs. (\ref{Zcosh}) and (\ref{Zsinh}) 
we can immediately read off the well-known asymptotic behavior of the 
orthogonal polynomials in the microscopic limit:
\beq
P_{2n}^{(0)}(\zeta) ~=~ C_1 \cos(\zeta) ~,~~~~~P_{2n+1}^{(0)}(\zeta)~=~
C_1 \sin(\zeta) ~.
\eeq
Although we needed to evaluate the partition function for an odd
number of fermions $N_f+1$ to obtain the polynomials for an even number $N_f$,
we have otherwise refrained from providing all the corresponding 
odd-$N_f$ results
for the double-microscopic spectral correlators, as well as the 
formulas for the orthogonal polynomials with an odd number of fermions. 
In fact, the physical interpretation
of the odd case is not as straightforward as that of the even case.
For example, in the massless limit the spectral density as normally
defined is only positive definite for even $N$ \cite{VZ}. Moreover, the
orthogonal polynomial technique does not directly apply, due to the
measure being odd under parity (in eigenvalue space) for odd $N$. Indeed,
even the formula (\ref{pol}) for the orthogonal polynomials will not be
valid in that case, due to the non-existence of normalizable orthogonal
polynomials. However, both partition functions (\ref{zeven})
and (\ref{zodd}), are completely well-defined, and can of course be used
for convenience to compute the orthogonal polynomials for {\em even}
$N_f$, as we have just shown.

\section{The Chiral Unitary Ensemble}

\noi
For the Dirac operator spectrum, this case corresponds to the 
gauge group $SU(N_c), N_c \geq 3$ with $N_f$ fermions in the fundamental 
representation. The matrix model
partition function reads, in the sector of topological charge 
$\nu$ (for convenience we shall consider $\nu\!\geq\!0$ throughout), \cite{V}
\beq
\tilde{\cal Z}_{\nu}^{(N_{f})}(m_1,\ldots,m_{N_{f}}) 
~=~ \int\! dW \prod_{f=1}^{N_{f}}{\det}\left(iM + m_f\right)~
\exp\left[-\frac{N}{2} \mbox{tr}\, V(M^2)\right] ~,\label{zMch}
\eeq
where
\beq
M ~=~ \left( \begin{array}{cc}
              0 & W^{\dagger} \\
              W & 0
              \end{array}
      \right) ~,
\eeq
where $W$ is a rectangular complex matrix of size
$N\times(N\! +\! \nu)$, which is integrated over with the Haar measure.
The space-time volume $V$ of the gauge theory is, in the large-$N$ (and
large-$V$) limit identified with $2N$. 

\noi
Introducing the eigenvalues $\la_i$ of the hermitian matrix $W^{\dagger}W$,
$\tilde{\cal Z}_{\nu}$ can be written 
\beq
\tilde{\cal Z}_{\nu}^{(N_{f})}(m_1,\ldots,m_{N_{f}})  ~=~ 
\prod_{f=1}^{N_{f}} 
(m_f^{\nu})\int_0^{\infty}\! \prod_{i=1}^N \left(d\lambda_i 
~\la_i^{\nu}~\prod_{f=1}^{N_{f}}(\lambda_i + m_f^2)~
\mbox{e}^{-NV(\lambda_i)}\right)\left|{\det}_{ij}
\lambda_j^{i-1}\right|^2 ~.\label{zmatrixeigen}
\eeq
We have ignored all unimportant factors that arise from the angular
integrations. Since the partition function for $N_{f}$ fermions
in the sector of topological charge $\nu$ is related in a simple way
to the partition function of the same $N_{f}$ fermions plus $\nu$ 
additional fermions of zero mass (in the sector of zero topological 
charge), one can in principle restrict attention to the $\nu\!=\! 0$ sector.
It is nevertheless worthwhile to point out that the whole analysis
which leads to a relation between the double-microscopic spectral
correlators and the finite-volume partition functions carries over to
the case of $\nu\!\neq\! 0$. This will lead us to very compact expressions
for these double-microscopic spectral correlators.

\noi
The necessary generalization of the previous analysis to the present case 
of a {\em chiral}
unitary ensemble with measure (\ref{zmatrixeigen}) is straightforward.
We are here interested in the spectral correlations of $M$-eigenvalues
$z_i$ rather than those of $W$-eigenvalues $\la_i\! =\! z_i^2$. Because
the whole procedure is identical to that of the previous section, and
because the case $\nu\!=\!0$ already has been worked out in detail \cite{D1},
we shall be brief. The two-point correlator, the kernel, is 
\begin{eqnarray}
&&K^{(N_{f},\nu)}_N(z,z';m_1,\ldots,m_{N_{f}}) ~=~ \cr
&& e^{-\frac{N}{2}(V(z^2)+V(z'^2))}(zz')^{\nu+\frac{1}{2}}
\prod_{f}^{N_{f}}\sqrt{(z^2+m_f^2)(z'^2+m_f^2)}
\sum_{i=0}^{N-1} P_i(z^2)P_i(z'^2) ~,
\end{eqnarray}
where $P_i(z^2)$ are the orthonormal polynomials associated with 
the above matrix model. The kernel can now be
expressed as a normalized random matrix integral:
\begin{eqnarray}
&&K^{(N_{f},\nu)}_N(z,z';m_1,\ldots,m_{N_{f}}) = 
\frac{e^{-\frac{N}{2}(V(z^2)+V(z'^2))}(zz')^{\nu+\frac{1}{2}}
\prod_{f}^{N_{f}}\sqrt{(z^2+m_f^2)(z'^2+m_f^2)}}{
\tilde{\cal Z}_{\nu}^{(N_{f})}(m_1,\ldots,m_{N_{f}})} \cr
&&~\times \prod_f^{N_f}(m_f^{\nu})
\int_0^{\infty}\! \prod_{i=1}^{N-1}\!\!\left(\!d\lambda_i 
\la_i^{\nu}(\la_i-z^2)(\la_i-z'^2)\!\!\prod_{f}^{N_{f}}(\lambda_i + m_f^2)
\mbox{e}^{-NV(\lambda_i)}\!\right)\!\left|{\det}_{ij}
\lambda_j^{i-1}\right|^2 ~.
\end{eqnarray}
Thus, in the large-$N$ limit we have
\begin{eqnarray}
K^{(N_{f},\nu)}_N(z,z';m_1,\ldots,m_{N_{f}}) &=&
e^{-\frac{N}{2}(V(z^2)+V(z'^2))}(-1)^{\nu}\sqrt{zz'}
\prod_{f}^{N_{f}}\sqrt{(z^2+m_f^2)(z'^2+m_f^2)}\cr
&&\times ~\frac{
\tilde{\cal Z}_{\nu}^{(N_{f}+2)}(m_1,\ldots,m_{N_{f}},iz,iz')}{
\tilde{\cal Z}_{\nu}^{(N_{f})}(m_1,\ldots,m_{N_{f}})} ~,
\end{eqnarray}
By means of the usual factorization property, all
higher $n$-point spectral correlation functions are then also explicitly
expressed in terms of the two matrix model partition functions
$\tilde{\cal Z}_{\nu}^{(N_{f})}$ and $\tilde{\cal Z}_{\nu}^{(N_{f}+2)}$.
The spectral density corresponds to the two additional (imaginary)
masses being equal, as in eq. (\ref{rhoUE}).

\noi
We now turn to the double-microscopic limit in which $\zeta \equiv
z N2\pi\rho(0)$ and $\mu_i \equiv m_i N2\pi\rho(0)$ are kept fixed as
$N\!\to\!\infty$. The prefactor 
$\exp[-(N/2)(V(z^2)+V(z'^2))]$ again becomes replaced by unity, and by
identifying $\Sigma = 2\pi\rho(0)$, we can now compare
with the field theory finite-volume partition functions.
This gives us the master formula 
\beq
K_S^{(N_{f},\nu)}(\zeta,\zeta';\mu_1,\ldots,\mu_{N_{f}}) ~=~ C_{2}
\sqrt{\zeta\zeta'}\prod_{f}^{N_{f}}
\sqrt{(\zeta^2+\mu_f^2)(\zeta'^2+\mu_f^2)}~\frac{
{\cal Z}_{\nu}^{(N_{f}+2)}(\mu_1,\ldots,\mu_{N_{f}},i\zeta,i\zeta')}{
{\cal Z}_{\nu}^{(N_{f})}(\mu_1,\ldots,\mu_{N_{f}})} ~.\label{mf}
\eeq
where the partition functions are those of the finite-volume field theories.
Similarly, for the double-microscopic spectral density,
\beq
\rho_S^{(N_{f},\nu)}(\zeta;\mu_1,\ldots,\mu_{N_{f}}) ~=~
C_2 |\zeta| \prod_{f}^{N_{f}}(\zeta^2+\mu_f^2)~\frac{
{\cal Z}_{\nu}^{(N_{f}+2)}(\mu_1,\ldots,\mu_{N_{f}},i\zeta,i\zeta)}{
{\cal Z}_{\nu}^{(N_{f})}(\mu_1,\ldots,\mu_{N_{f}})} ~.
\eeq
All double-microscopic $n$-point correlation functions are again given
by the factorization formula (\ref{correlUE}).

\noi
A simple example which illustrates how powerful the above relations can be
is that of the quenched case $N_f\!=\!0$. All we need is the
finite-volume QCD partition function for two massive fermions of
degenerate masses $i\mu/(N\Sigma)$. This was evaluated analytically
already by Leutwyler and Smilga \cite{LS} and found to be, 
in their normalization,
\beq
{\cal Z}_{\nu}^{(2)}(i\mu,i\mu) ~=~ I_{\nu}(i\mu)^2 - 
I_{\nu+1}(i\mu)I_{\nu-1}(i\mu) ~,
\eeq
where $I_n(x)$ is the $n$th modified Bessel function. The corresponding
denominator in eq. (\ref{mf}) is again
an irrelevant constant which we can set to unity. This gives
\beq
\rho_S^{(0,\nu)}(\zeta) = C_2 (-1)^\nu|\zeta|\left[J_{\nu}(\zeta)^2
- J_{\nu+1}(\zeta)J_{\nu-1}(\zeta)\right] ~.
\eeq
The matching condition (\ref{match}) yields
$$
C_2 = \frac{1}{2}(-1)^{\nu} ~,
$$
and hence
\beq
\rho_S^{(0,\nu)}(\zeta) = \frac{1}{2}|\zeta|\left[J_{\nu}(\zeta)^2
- J_{\nu+1}(\zeta)J_{\nu-1}(\zeta)\right] ~,
\eeq
which is the known result \cite{V}. Furthermore, by the previous
considerations (cf. eq. (\ref{zmatrixeigen})) we also know that the
general case of $N_f$ massless fermions simply is equivalent to a shift
$\nu\!\to\! \nu\!+\!N_f$. We thus recover the general massless result
\cite{V} without any effort:
\beq
\rho_S^{(N_{f},\nu)}(\zeta) ~=~ \frac{1}{2}|\zeta|\left[J_{N_{f}+\nu}(\zeta)^2
- J_{N_f+\nu+1}(\zeta)J_{N_f+\nu-1}(\zeta)\right] ~.
\eeq

\noi
We now need the general analytical expression for the finite-volume partition
function for 
this case, with $N_f$ fermions of arbitrary masses 
\cite{JSV} (see also ref. \cite{Vrev}). It can conveniently be written
\cite{D1}  
\beq
{\cal Z}_{\nu}^{(N_{f})}(\mu_1,\ldots,\mu_{N_{f}}) ~=~ 
\frac{\det {\cal A}(\{\mu_i\})}{\Delta(\mu^2)} 
\eeq
where the $N_f\!\times\! N_f$ matrix ${\cal A}(\{\mu_i\})$ is
\beq
{\cal A}_{ij} ~=~ \mu_i^{j-1}I_{\nu+j-1}(\mu_i) ~,\label{Aconv}
\eeq
and $\Delta(\mu^2)$ again indicates the 
Vandermonde determinant of the $\mu_i^2$.

\noi
For the numerator of eq. ({\ref{mf}) we need the $(N_f\!+\!2)\!\times\!
(N_f\!+\!2)$ matrix ${\cal A}$ with two of the entries being imaginary. 
This means that
\beq
{\cal A}_{ij} ~=~ i^{\nu}(-\zeta_i)^{j-1}J_{\nu+j-1}(\zeta_i) ~~~~~~~~~~ 
{\mbox{\rm for}}~~i = 1,2 ~,
\eeq
and otherwise (for $i\!\geq\!3$) as in eq. (\ref{Aconv}). 
For convenience we pull out a
factor of (-1) from every second column of the matrix ${\cal A}$, and also
the factor of $i^{\nu}$ from the first two rows. This yields an
overall factor of $(-1)^{\nu + [N_f/2]}$
where $[x]$ denotes the integer part of $x$. 
Thus,
\beq
K_S^{(N_{f},\nu)}(\zeta_1,\zeta_2;\mu_1,\ldots,\mu_{N_{f}}) =
C_2~\frac{(-1)^{\nu+[N_f/2]+1}\sqrt{\zeta_1\zeta_2}}{(\zeta_1^2 - \zeta_2^2)
\prod_f\sqrt{(\zeta_1^2+\mu_f^2)(\zeta_2^2+\mu_f^2)}}~
\frac{\det {\cal B}}{\det {\cal A}} ~,
\label{KS}
\eeq
where the $(N_f\!+\!2)\!\times\!(N_f\!+\!2)$ matrix ${\cal B}$ is defined by
\begin{eqnarray}
{\cal B}_{ij} &=& (\zeta_i)^{j-1}J_{\nu+j-1}(\zeta_i) ~~~~~~~~~~~~~~~~
{\mbox{\rm for}}~~i = 1,2 \cr
{\cal B}_{ij} &=& (-\mu_{i-2})^{j-1}I_{\nu+j-1}(\mu_{i-2}) ~~~~~~~
{\mbox{\rm for}}~~ 3 \leq i \leq N_f+2 ~,
\end{eqnarray}
and the $N_f\!\times\! N_f$ matrix ${\cal A}$ is as in (\ref{Aconv}). Using
the Bessel relation
\beq
\frac{d}{dx}\left[x^nJ_{n+m}(x)\right] ~=~ x^nJ_{n+m-1}(x) - mx^{n-1}
J_{n+m}(x)
\eeq
we find the corresponding double-microscopic spectral density:
\beq
\rho_S^{(N_{f},\nu)}(\zeta;\mu_1,\ldots,\mu_{N_{f}}) ~=~
C_2~\frac{(-1)^{\nu+[N_f/2]+1}|\zeta|}{2
\prod_f(\zeta^2+\mu_f^2)}~\frac{\det \tilde{{\cal B}}}
{\det {\cal A}} ~,\label{rhoS}
\eeq
where the  $(N_f\!+\!2)\!\times\!(N_f\!+\!2)$ matrix $\tilde{\cal B}$ 
is defined by
\beq
\tilde{{\cal B}}_{1j} ~=~ \zeta^{j-2}J_{\nu+j-2}(\zeta) 
\eeq
and $\tilde{{\cal B}}_{ij}\!=\! {\cal B}_{ij}$ for $i\!\neq\! 1$. 
The general $n$-point
correlators follow from eqs. (\ref{correlUE}) and (\ref{KS}). Using the
matching condition (\ref{match}) gives 
$$
C_2 ~=~ (-1)^{\nu+[N_f/2]} ~,
$$
and everything is now determined.

\noi
For $\nu\!=\!0$ the results (\ref{KS}) and (\ref{rhoS}) agree with what 
has recently been obtained by a
direct computation in random matrix theory \cite{DN,WGW}. While no
explicit expressions were given for the case $\nu\!\neq\! 0$ in ref. \cite{DN},
it could in principle be extracted from the general formulae for $\nu\!=\!0$ 
by setting $\nu$ fermion masses equal to zero in a theory of $N_f\!+\!
\nu$ fermions. The result done in that way should of course agree with
our explicit formula given above. Indeed, we have managed to prove by 
induction that the compact formulas (\ref{KS}) and (\ref{rhoS}) also follow
from the $\nu\!=\!0$ results given in ref. \cite{DN}.

\noi
As in the previous section for the 
unitary ensemble, we now show that also
in the chiral case the universal limit of the orthogonal polynomials can be
obtained directly from the finite-volume partition functions
alone. The extension of the formula (\ref{polz}) to this chiral
case is straightforward. There is now no parity ``quantum number'' for
the polynomials, and the universal double-microscopic limit is therefore
unique, independent of whether the polynomials are of odd or even order.
The explicit formula reads
\beq
P_{n}^{(N_{f},\nu)}(\la;m_1,\ldots,m_{N_{f}}) ~=~ 
\frac{\prod_f^{N_f}(m_f^\nu)}
{\tilde{\cal Z}^{N_f}_{\nu,n}(m_1,\ldots,m_{N_{f}})}\int_{0}^{\infty}
\prod_{i=1}^{n}[d\la_i w(\la_i)]\left|{\det}_{ij}\lambda_j^{i-1}\right|^2 
\prod_{i=1}^{n}(\la - \la_i) ~,\label{polch}
\eeq
where now the weight function $w(\la)$ is given by
\beq
w(\la) ~=~ \la^{\nu}\prod_{f=1}^{N_f}\left[(\la + m_f^2)\right]
~\mbox{e}^{-NV(\la)} ~,\label{wch}
\eeq
and
\beq
\tilde{\cal Z}^{(N_f)}_{\nu,n} ~=~ \prod_f^{N_f}(m_f^\nu)
\int_{0}^{\infty}\prod_i^{n}[d\la_i w(\la_i)]
\left|{\det}_{ij}\lambda_j^{i-1}\right|^2 ~,
\eeq
Also here one easily sees that the normalization is such that the polynomials 
$P_{n}^{(N_{f},\nu)}(\la;m_1,\ldots,m_{N_{f}})$ are monic.

\noi
In the limit $n\!\to\!\infty$ and $N\!\to\!\infty$ with $t=n/N$ fixed,
the relation (\ref{polch}) determines the orthogonal polynomials in
terms of the matrix model partition functions. By going to the
double-microscopic scaling regime with $t=1$ where we can make the 
identification with the field theory partition functions, this gives us 
the relation
\beq
P_{N}^{(N_{f},\nu)}(\zeta^2;\mu_1,\ldots,\mu_{N_{f}}) ~=~ C_3 
(-1)^N(i\zeta)^{-\nu}\
\frac{{\cal Z}^{(N_{f}+1)}_\nu(\mu_1,\ldots,\mu_{N_{f}},i\zeta)}
{{\cal Z}^{(N_{f})}_\nu(\mu_1,\ldots,\mu_{N_{f}})} ~, \label{polzch}
\eeq
where the normalization constant $C_3$ is still undetermined and we have
passed to scaled $M$-eigenvalues. This relation
is just as in the ordinary unitary case. For the numerator we need the
$(N_f+1)\times(N_f+1)$ matrix ${\cal A}$ eq. (\ref{Aconv}) with one imaginary 
entry, 
\beq
{\cal A}_{1j} ~=~ i^\nu (-\zeta)^{j-1}J_{\nu+j-1}(\zeta)
\eeq
and otherwise as in eq. (\ref{Aconv}). In order to fix the constant $C_3$
and to compare with \cite{DN} we choose the normalization 
$P_{N}^{(N_{f},\nu)}(0;\mu_1,\ldots,\mu_{N_{f}})\!=\!1$.
We then obtain 
\beq
P_{N}^{(N_{f},\nu)}(\zeta^2;\mu_1,\ldots,\mu_{N_{f}}) ~=~
\zeta^{-\nu}\prod^{N_f}_f\frac{\mu_f^2}{\zeta^2+\mu_f^2}
\frac{\det{\cal D}}{\det\tilde{\cal A}} 
\label{OPUE}
\eeq
where 
\beqn
{\cal D}_{1j}&=& (-\zeta)^{j-1}J_{\nu+j-1}(\zeta) ~~~~~~~~~~~~
\mbox{for}~~ 1\leq j\leq N_f+1 ~~,\nonumber\\
{\cal D}_{ij}&=& \mu_{i-1}^{j-1}I_{\nu+j-1}(\mu_{i-1}) ~~~~~~~~~~~~~
\mbox{for}~~ 2\leq i\leq N_f+1 ~;~~~~1\leq j\leq N_f+1 ~,
\eeqn
and
\beq
\tilde{\cal A}_{ij} ~=~ \mu_i^j I_{\nu+j}(\mu_i) ~~~~~~~
\mbox{for}~~ 1\leq i,j\leq N_f ~.
\eeq
For $\nu\!=\! 0$ the above expression matches to the result for the 
polynomials in \cite{DN}, which was derived from random matrix theory. 
Here, eq. (\ref{OPUE}) directly gives the results for $\nu$ zero modes,
which is equivalent to set $\nu$ fermion masses to 0 in \cite{DN}. We have
proven by induction that the results of ref. \cite{DN} lead to precisely
the same formula for arbitrary $\nu$ as shown above.

\section{The Symplectic and Orthogonal Ensembles}

As follows from the general classification of universality classes \cite{V},
gauge group $SU(2)$ and $N_f$ fermions in the fundamental representation
correspond in matrix model language to the orthogonal ensemble, while
gauge group $SU(N_c)$ with $N_f$ fermions in the adjoint representation
of the gauge group correspond to the symplectic ensemble.

\noi
The symplectic and orthogonal matrix ensembles are somewhat more
complicated from an analytical point of view due to the non-existence of 
simple orthogonal-polynomial methods for those cases. The closest one
apparently can get is based on the so-called quaternion method, which
can be phrased in terms of skew-orthogonal (as opposed to truly
orthogonal) polynomials \cite{MM}. 
In this chapter we shall consider what may be the
closest symplectic and orthogonal ensemble analogues of the relations 
derived above for the unitary and chiral unitary ensembles.

\noi
We begin with the chiral symplectic matrix ensemble, as this is the
case for which we most easily can derive useful relations that connect
finite-volume partition functions to the associated Dirac spectra.
In the language of random matrix theory, the partition function
for the chiral symplectic ensemble is as in eq. (\ref{zMch}), except
that now the integration is over matrices $W$ whose elements are
quaternion real \cite{V}.
In terms of the eigenvalues $\la_i$ of the hermitian matrix 
$W^{\dagger}W$, $\tilde{\cal Z}_{\nu}$ can now be written (we follow
the conventional normalization, where the symplectic matrix model potential
is rescaled by a factor of 2 compared with the chiral unitary case):
\beq
\tilde{\cal Z}_{\nu}^{(N_{f})}(m_1,\ldots,m_{N_{f}})  ~=~ 
\prod_{f=1}^{N_{f}}(m_f^{\nu})
\int_0^{\infty}\! \prod_{i=1}^N \left(d\lambda_i 
~\la_i^{2\nu+1}~\prod_{f=1}^{N_{f}}(\lambda_i + m_f^2)~
\mbox{e}^{-2NV(\lambda_i)}\right)\left|{\det}_{ij}
\lambda_j^{i-1}\right|^4 ~.\label{zsymmatrixeigen}
\eeq

\noi
Our goal is now to find the closest analogues of the master formulas
(\ref{mfUE}) and (\ref{mf}). We shall make good use of some general relations
that have been derived by Mahoux and Mehta \cite{MM}. Throughout this
section we will use their notation here.

\noi
The problem we encounter is that the quantity that most closely corresponds 
to the kernel of the now skew-orthogonal polynomials now is a
{\it quaternion} $f_4(\la_i,\la_j)$, which can be represented by
a $2\!\times\!2$ matrix. The correlation functions of eigenvalues are then
given by quaternion determinants $\det[f_4(\la_i,\la_j)]_m$ of the kernel
$f_4(\la_i,\la_j)$. We have not been able to express this kernel itself
in terms of matrix model (and thus also finite-volume field theory)
partition functions, but only the determinants of this kernel, which are real
valued functions.
This will directly give us the expressions for the correlators, where we
display the eigenvalue density and the density-density correlator as examples.
Going back to eigenvalues $z_i$ of the Dirac operator rather than
$\la_i=z_i^2$ the spectral density can be obtained in the following way:
\beqn
\rho_4^{(N_{f},\nu)} (z;m_1,\ldots,m_{N_{f}}) 
&=& \frac{1}{N}\det[f_4(z,z)]_1 \cr
&=& z^{4\nu+3}\prod_{f=1}^{N_f}(z^2+m_f^2)~
\mbox{e}^{-2NV(z^2)}\frac{\prod_{f=1}^{N_{f}}(m_f^{\nu})}
{\tilde{\cal Z}_{\nu}^{(N_{f})}(m_1,\ldots,m_{N_{f}})} \cr
&\times &
\int_0^{\infty}\prod_{i=1}^{N-1}\left( dz_i~ z_i^{4\nu+3}
~\prod_{f=1}^{N_{f}}(z_i^2 + m_f^2)~
\mbox{e}^{-2NV(z_i^2)} |z^2-z_i^2|^4\right)\left|{\det}_{ij}
z_j^{2i-2}\right|^4 \cr
&=& i^{-4\nu}z^3\prod_{f=1}^{N_f}(z^2+m_f^2)~\mbox{e}^{-2NV(z^2)}
\frac{\tilde{\cal Z}_{\nu}^{(N_{f}+4)}
(m_1,\ldots,m_{N_{f}},\{iz\})}
{\tilde{\cal Z}_{\nu}^{(N_{f})}(m_1,\ldots,m_{N_{f}})}
~.\label{rho4}
\eeqn
In the first step we have made use of Theorem 1.2
of ref. \cite{MM} in the form
\beq
\int dz_{p+1}\det[f_4(z_i,z_j)]_{p+1} ~=~ 
(N-p)\det[f_4(z_i,z_j)]_p ~~,
\label{theo1.2}
\eeq
using their explicit expression for the partition function. 
We have slightly generalized the measure of ref. \cite{MM} here to
include the massive fermions and zeromodes. In the second step
we have replaced the integral over $N\!-\!1$ eigenvalues by the matrix model
partition function with 4 equal massive flavors of imaginary mass 
$iz$, ignoring their difference in the large-$N$ limit.

\noi
When replacing the matrix model partition function in the scaling limit
by its mesoscopic field theory counterpart, 
as was done in the previous sections, we obtain the following relation:
\beq
\rho_S^{(N_{f},\nu)}(\zeta;\mu_1,\ldots,\mu_{N_{f}}) ~=~ C_4~ \zeta^3
\prod_{f=1}^{N_f}(\zeta^2+\mu_f^2)~
\frac{{\cal Z}^{(N_{f}+4)}_\nu(\mu_1,\ldots,\mu_{N_{f}},\{i\zeta\})}
{{\cal Z}^{(N_{f})}_\nu(\mu_1,\ldots,\mu_{N_{f}})} ~.
\label{symquenched}
\eeq
Unfortunately the needed finite-volume partition function of 4 or more 
fermions in the adjoint
representation is not known at present and thus the eigenvalue density 
cannot be further evaluated yet. But since from matrix model calculations the 
eigenvalue density is known
to be expressible in terms of integrals of Bessel functions \cite{NF} 
(see also \cite{BBMSVW})
a relatively simple expression for the field theory partitions functions 
should exist. (Simple analytical formulas are indeed known at present up 
to $N_f\!=\!2$ \cite{LS,D}).

\noi
As a second example we derive in a similar way an expression for the 
density-density correlator:
\beqn
\rho_4^{(N_f,\nu)}(z,z';\{m_f\}) &=& \frac{1}{N(N-1)}\det[f_4(z,z')]_2 \cr
&=& |z^2-{z'}^2|^4(zz')^{4\nu+3}\prod_{f=1}^{N_f}(z^2+m_f^2)(z'^2+m_f^2)~
\mbox{e}^{-2N(V(z^2)+V({z'}^2))}
\frac{\prod_{f=1}^{N_{f}}(m_f^{\nu})}
{\tilde{\cal Z}_{\nu}^{(N_{f})}(\{m_f\})} \cr
&\times&\int_0^{\infty}\prod_{i=1}^{N-2}\left( dz_i z_i^{4\nu+3}
\prod_{f=1}^{N_{f}}(z_i^2 + m_f^2)~\mbox{e}^{-2NV(z_i^2)} 
|z^2-z_i^2|^4|{z'}^2-z_i^2|^4\right)\left|{\det}_{ij}
z_j^{2i-2}\right|^4 \cr
&=& (-1)^{4\nu}|z^2-{z'}^2|^4(zz')^3\prod_{f=1}^{N_f}(z^2+m_f^2)(z'^2+m_f^2)~
\mbox{e}^{-2N(V(z^2)+V({z'}^2))} \cr
&\times&
\frac{\tilde{\cal Z}_{\nu}^{(N_f+4+4)}(m_1,\ldots,m_{N_f},\{iz\},\{iz'\})}
{\tilde{\cal Z}_{\nu}^{(N_f)}(m_1,\ldots,m_{N_f})} ~~~.
\eeqn 
and thus in the double-microscopic limit,
\beq
\rho_S^{(N_{f},\nu)}(\zeta,\zeta';\mu_1,\ldots,\mu_{N_{f}}) = 
\tilde{C_4} |\zeta^2-\zeta'^2|^4 (\zeta\zeta')^3 
\prod_{f=1}^{N_f}(\zeta^2+\mu_f^2)(\zeta'^2+\mu_f^2)
\frac{{\cal Z}_{\nu}^{(N_{f}+8)}(\mu_1,\ldots,\mu_{N_{f}},
\{i\zeta\},\{i\zeta'\})}
{{\cal Z}_{\nu}^{(N_{f})}(\mu_1,\ldots,\mu_{N_{f}})}
\eeq

\noi
In the unitary ensembles the knowledge of this two-point
correlator is equivalent to knowing the kernel as well, 
since its connected part
is the square of the kernel $\rho_{con}(z,z')\!=\!-\!K(z,z')^2$.
However, in the symplectic case the kernel $f_4(z_i,z_j)$ is not
a symmetric function and does not factorize. All higher 
correlation functions can actually be expressed by partition functions 
in an analogous way as above.

\noi
Finally we turn to the case of the orthogonal ensemble, for which the
relevant partition function, when expressed in terms of eigenvalue
integrals, reads
\beq
\tilde{\cal Z}_{\nu}^{(N_{f})}(m_1,\ldots,m_{N_{f}})  ~=~ 
\prod_{f=1}^{N_{f}} 
(m_f^{\nu})\int_0^{\infty}\! \prod_{i=1}^N \left(d\lambda_i 
~\la_i^{\frac{\nu}{2}-\frac{1}{2}}~\prod_{f=1}^{N_{f}}(\lambda_i + m_f^2)~
\mbox{e}^{-\frac{N}{2}V(\lambda_i)}\right)\left|{\det}_{ij}
\lambda_j^{i-1}\right| ~.\label{zorthmatrixeigen}
\eeq
Since the orthogonal case can only be treated on the same footing as the 
symplectic case using quaternions \cite{MM}, we will follow the same procedure
as above:
\beqn
\rho_1^{(N_f,\nu)} (z;\{m_i\}) &=& \frac{1}{N}\det[f_1(z,z)]_1 \cr
&=& z^{\nu}\prod_{f=1}^{N_f}(z^2+m_f^2)~\mbox{e}^{-\frac{N}{2}V(z^2)}
\frac{\prod_{f=1}^{N_{f}}(m_f^{\nu})}
{\tilde{\cal Z}_{\nu}^{(N_{f})}(m_1,\ldots,m_{N_{f}})}\cr
&\times& 
\int_0^{\infty}\prod_{i=1}^{N-1}\left( dz_i~ z_i^{\nu}
\prod_{f=1}^{N_{f}}(z_i^2 + m_f^2)~\mbox{e}^{-\frac{N}{2}V(z_i^2)} 
|z^2-z^2_i|\right)\left|{\det}_{ij} z_j^{2i-2}\right|
\label{rho1}
\eeqn
and
\beqn
\rho_1^{(N_f,\nu)}(z,z';\{m_i\}) &=& \frac{1}{N(N-1)}\det[f_1(z,z')]_2 \cr
&=& |z^2-{z'}^2|(zz')^{\nu}\prod_{f=1}^{N_f}(z^2+m_f^2)(z'^2+m_f^2)
\mbox{e}^{-\frac{N}{2}(V(z^2)+V({z'}^2))}\frac{\prod_{f=1}^{N_{f}}(m_f^{\nu})}
{\tilde{\cal Z}_{\nu}^{(N_{f})}(m_1,\ldots,m_{N_{f}})} \cr
&\times&\int_0^{\infty}\prod_{i=1}^{N-2}\left( dz_i~ z_i^{\nu}
\prod_{f=1}^{N_{f}}(z^2_i + m_f^2)~\mbox{e}^{-\frac{N}{2}V(z^2_i)} 
|z^2-z^2_i||{z'}^2-z_i^2|\right)\left|{\det}_{ij}
z_j^{2i-2}\right| .
\label{corr1}
\eeqn
It is straightforward to go to the double-microscopic limit of these
expressions.
However, in this case the absolute value inside the eigenvalue integrals 
eqs. (\ref{rho1}) and (\ref{corr1}) 
prevents us from immediately identifying them with the matrix model 
partition function with
additional masses in any simple way. 
But since analogously to the symplectic ensemble 
an expression for the microscopic density in terms of integrals
of Bessel functions is known \cite{V}, similar relations 
to finite-volume partition functions should exist.

\noi
In this context it is particularly interesting to consider those relations
between the kernels of the chiral symplectic/orthogonal ensembles and
the chiral unitary ensemble which very recently have been derived by
\c{S}ener and Verbaarschot \cite{cSV}. According to our present viewpoint,
these relations not only extend the universality proof of ref. \cite{ADMN}
to these ensembles, but also provide completely surprising and non-trivial
identities among finite-volume partition functions for different effective
field theories. An understanding of these new relations between effective
partition functions would be very desirable at this point.

\section{Conclusions}

\noi
In this paper we have systematically explored the relationship between
spectral correlators of the Dirac operator in the double-microscopic
scaling region, and the corresponding finite-volume partition functions.
Based on relations that can be proven in random matrix theory, we have
shown how to extract the universal properties from these finite-volume
partition functions alone. The most powerful results are the two master
formulas for the unitary and chiral unitary ensembles. These allow for
a complete determination of all double-microscopic spectral correlators
for those cases in terms of finite-volume partition functions. 
One of the surprising conclusions
is that one can also derive the
universal limits of the orthogonal polynomials of random matrix theory
from the associated field theory partition functions.

\noi
The resulting relations are therefore not just of interest from the
field theory point of view (where they indicate that
the formulation in terms of large-$N$ random matrix theory to some extent
can be avoided), but also from the viewpoint of random matrix theory.
Indeed, in section 2 we have illustrated this in another way by showing
how the famous (bulk) sine kernel of the unitary ensemble can be 
derived neatly from a simple $SU(2)$ ``chiral lagrangian''. Many other examples
can surely be found.   

\noi
The cases most exhaustively solved are those of the unitary and chiral unitary
matrix ensembles. Here all pertinent information for the orthogonal 
polynomials, their kernel and thus all correlation functions are derivable
from the corresponding finite-volume partitions, suitably extended to
include more fermionic species of imaginary masses. 
For the chiral symplectic random matrix ensemble we have chosen a different 
way. We have directly related the correlation functions
to the corresponding field theory partition functions with 4-fold 
degenerate additional fermion species.
These relations may turn out to provide the most easy analytical derivation
of the involved quantities. For the chiral orthogonal case there is
a highly suggestive relation, which, however, relates the spectral
density to the partition function of a theory where the absolute value
of the determinant of the Dirac operator enters. More work is required
here, to either relate this partition function to the conventional one, 
or to construct a chiral lagrangian that would correspond to taking
the absolute value of the Dirac determinant. 

\noi
It is also worthwhile to reconsider the notion of random matrix 
universality in this new light. Of course, there is no substitute for
the complete mathematical proof \cite{ADMN}. But we can gain substantial
insight into the mechanism of universality by tracing the disappearance
of the matrix model potential $V(\la)$ in the relevant expressions.
If we, for example, return to eq. (\ref{KtildeZUE}) of random matrix theory
and the master formula (\ref{mfUE}), we see that in the double-microscopic
scaling limit the prefactor of $\exp[-(N/2)(V(x^2)+V(x'^2))]$ simply
becomes replaced by unity. This is the only place where the random matrix
theory potential enters {\em explicitly} (and where it disappears in
the scaling limit, leaving a universal result). Corresponding factors
of the potentials disappear in the other analogous relations. 
One should not be misled by these simple observations to conclude that
universality of the matrix model results can be understood in such simple
terms alone. To some extent the notion of universality is simply built into
the crucial identification between matrix model and field theory
partition functions in the mesoscopic, or double-microscopic, scaling
regime. Indeed, the disappearance of factors such as
$\exp[-(N/2)(V(x^2)+V(x'^2))]$ in the appropriate scaling limit is not
the sole mechanism behind the proven universality of random matrix theory
results. An obvious counterexample is provided by the recent study
of microscopic limits of random matrix theories for which the 
macroscopic spectral density $\rho(0)$ at the origin precisely is
vanishing \cite{ADMN1}. Here exponential prefactors such as those 
discussed above do {\em not} approach unity in the microscopic scaling limits,
and still universality can be proven \cite{ADMN1}. It would be
interesting to find the chiral lagrangian analogues of these multicritical
cases (for which $\rho(0)\!=\! 0$\footnote{There are hence no Goldstone modes,
and no obvious group manifold on which to base the effective Lagrangian of
the lightest hadronic excitations. A new principle seems needed to derive
the corresponding effective Lagrangian.}), 
and explore the relations discussed here in that more general context.  

\noi
{\sc Acknowledgment:}\\
 The work of G.A. is supported by European Community
grant no. ERBFMBICT960997 and the work of P.H.D. is partially supported by
EU TMR grant no. ERBFMRXCT97-0122.


\end{document}